\documentclass[a4paper,11pt]{article}
\pdfoutput=1 % if your are submitting a pdflatex (i.e. if you have
\usepackage{jheppub} % for details on the use of the package, please
\usepackage[T1]{fontenc} % if needed
\usepackage{physics}
\usepackage{dsfont}
\usepackage{bm}

\usepackage{tikz}
\usetikzlibrary{arrows,shapes}
\usetikzlibrary{trees}
\usetikzlibrary{matrix,arrows} 				% For commutative diagram
\usetikzlibrary{positioning}				% For "above of=" commands
\usetikzlibrary{calc,through}				% For coordinates
\usetikzlibrary{decorations.pathreplacing}  % For curly braces
\usepackage{pgffor}							% For repeating patterns
\usetikzlibrary{decorations.pathmorphing}	% For Feynman Diagrams
\usetikzlibrary{decorations.markings}
\tikzset{
	vector/.style={decorate, decoration={snake}, draw},
	provector/.style={decorate, decoration={snake,amplitude=2.5pt}, draw},
	antivector/.style={decorate, decoration={snake,amplitude=-2.5pt}, draw},
	fermion/.style={draw=black, postaction={decorate},
	decoration={markings,mark=at position .55 with {\arrow[draw=black]{>}}}},
	fermionbar/.style={draw=black, postaction={decorate},
	decoration={markings,mark=at position .55 with {\arrow[draw=black]{<}}}},
	fermionnoarrow/.style={draw=black},
	gluon/.style={decorate, draw=black,
	decoration={coil,amplitude=4pt, segment length=5pt}},
	scalar/.style={dashed,draw=black, postaction={decorate},
	decoration={markings,mark=at position .55 with {\arrow[draw=black]{>}}}},
	scalarbar/.style={dashed,draw=black, postaction={decorate},
	decoration={markings,mark=at position .55 with {\arrow[draw=black]{<}}}},
	scalarnoarrow/.style={dashed,draw=black},
	electron/.style={draw=black, postaction={decorate},
	decoration={markings,mark=at position .55 with {\arrow[draw=black]{>}}}},
	bigvector/.style={decorate, decoration={snake,amplitude=4pt}, draw},
}

\def\dsH{\mathds{H}}
\def\dsM{\mathds{M}}
\def\dsGamma{\text{I}\hspace{-2.25pt}\Gamma}

\title{Stimulated transitions in resonant atom Majorana mixing}

\author{Jos\'e Bernab\'eu}
\author{and Alejandro Segarra}
\affiliation{Departament de F\'isica Te\`orica and IFIC, Universitat de Val\`encia - CSIC, E-46100, Spain}

\emailAdd{Jose.Bernabeu@uv.es}
\emailAdd{Alejandro.Segarra@uv.es}

\abstract{Massive neutrinos demand to ask whether they are Dirac or Majorana particles. 
Majorana neutrinos are an irrefutable proof of physics beyond the Standard Model. 
Neutrinoless double electron capture is not a process but a virtual $\Delta L = 2$ mixing 
between a parent $^AZ$ atom and a daughter $^A(Z-2)$ excited atom with two electron holes. 
As a mixing between two neutral atoms and the observable signal in terms of emitted two-hole X-rays, 
the strategy, experimental signature and background are different from neutrinoless double beta decay. 
The mixing is resonantly enhanced for almost degeneracy and, under these conditions, 
there is no irreducible background from the standard two-neutrino channel. 
We reconstruct the natural time history of a nominally stable parent atom 
since its production either by nature or in the laboratory. 
After the time periods of atom oscillations and the decay of the short-lived daughter atom, 
at observable times the relevant ``stationary''  states are the mixed metastable long-lived state 
and the non-orthogonal short-lived excited state, as well as the ground state of the daughter atom. 
We find that they have a
natural population inversion which is most appropriate for exploiting the bosonic nature 
of the observed atomic transitions radiation. 
Among different observables of the atom Majorana mixing, 
we include the enhanced rate of stimulated X-ray emission from the long-lived 
metastable state by a high-intensity X-ray beam:
a gain factor of 100 can be envisaged 
at current XFEL facilities.
On the other hand, the historical population of the daughter atom ground state
can be probed by exciting it with a current pulsed optical laser, showing the
characteristic absorption lines: the whole population can be excited in a
shorter time than typical pulse duration.}

\begin{document} 
\maketitle
\flushbottom

\section{Introduction}

The experimental evidence of neutrino oscillations is one of the most important discoveries in particle physics. Model-independent first evidences of neutrino oscillations were obtained in 1998 by the atmospheric neutrino experiment Super-Kamiokande \cite{1}, in 2002 by the solar neutrino experiment SNO \cite{2}, and later by accelerator and reactor neutrinos.

The existence of neutrino oscillations implies that neutrinos are massive particles and that the three flavor neutrinos $\nu_e,\nu_\mu, \nu_\tau$ are mixtures of the neutrinos with definite masses $\nu_i$ (with i = 1, 2, 3). The phenomenon of neutrino oscillations is being studied in a variety of experiments \cite{3} which fully confirm this quantum phenomenon in different disappearance and appearance channels. The mixing matrix $U_\text{PMNS}$ contains three mixing angles, already known, and one CP violating phase for flavor oscillations. Interacting neutrinos have left-handed chirality.

Knowing that neutrinos are massive, the most fundamental open problem is the determination of the nature of neutrinos with definite mass: are they four-component Dirac particles with a conserved total lepton number $L$, distinguishing neutrinos from antineutrinos, or two-component truly neutral (no electric charge and no total lepton number) self-conjugate Majorana particles \cite{4}? For Dirac neutrinos, like quarks and charged leptons, their masses can be generated in the Standard Model of particle physics by spontaneous breaking of the gauge symmetry with the Higgs scalar, if there were additional right-handed sterile neutrinos. But the Yukawa couplings would then be unnaturally small compared with all other fermions. A Majorana $\Delta L = 2$ mass term, with the active left-handed neutrinos only, leads to definite mass neutrinos with no definite charge. However, there is no way in the Standard Model able to generate this Majorana mass, so the important conclusion in fundamental physics arises: Majorana neutrinos are an irrefutable proof of physics beyond the Standard Model. Due to the Majorana condition of neutrinos with definite mass being their own antiparticles, Majorana neutrinos have additional CP violating phases \cite{5,6,7} beyond the Dirac case.

Neutrino flavor oscillation experiments cannot answer the fundamental question of the nature of massive neutrinos, because in these flavor transitions the total lepton number $L$ is conserved. In order to probe whether neutrinos are Dirac or Majorana particles, we need to study observables violating the total lepton number $L$. The difficulty encountered in these studies is well illustrated by the so-called ``confusion theorem" \cite{8,9}, stating that in the limit of zero-mass there is no difference between Dirac and Majorana neutrinos. As all known neutrino sources produce highly relativistic neutrinos (except for the present cosmic neutrino background in the universe), the $\Delta L = 2$ observables are highly suppressed. 

Up to now, there is a consensus that the highest known sensitivity to small Majorana neutrino masses can be reached in experiments on the search of the $L$-violating neutrinoless double-$\beta$ decay process ($0\nu\beta\beta$)
\begin{equation}
	\label{eq:0nbb}
           ^AZ \to\, ^A(Z+2) + 2e^-\,,
\end{equation}
where $^AZ$ is a nucleus with atomic number $Z$ and mass number $A$. The two-neutrino double-$\beta$ decay process ($2\nu\beta\beta$):
\begin{equation}
	\label{eq:2nbb}
	^AZ \to\, ^A(Z+2) + 2 e^- + 2 \bar \nu_e
\end{equation}
is allowed by the Standard Model for some even-even nuclei for which the single $\beta$ decay or electron capture is forbidden. The process (\ref{eq:2nbb}) represents an irreducible background in the search of (\ref{eq:0nbb}), which needs an excellent energy resolution in order to separate the definite peak in (\ref{eq:0nbb}) from the high energy tail of the $2e^-$ spectrum in (\ref{eq:2nbb}).

\begin{figure}[t]
	\centering
	\includegraphics[width=0.6\textwidth]{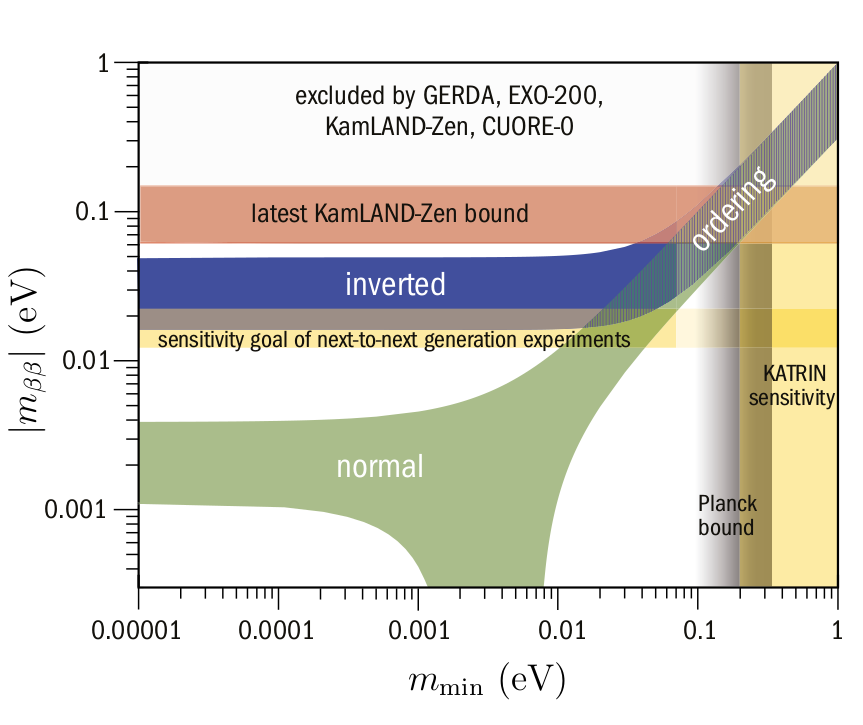}
	\caption{Allowed parameter space for a $0\nu\beta\beta$ signal, as a function of the smallest neutrino mass, for both possible hierarchies.
	Current experimental bounds are also shown. Taken from \cite{pascoli}.}
	\label{fig:1}
\end{figure}

Dozens of experiments around the world are seeking out a positive signal of $0\nu\beta\beta$. The most favorable decays for the experimental search are those with high mass difference between the ground state neutral atoms. The most sensitive limits at present are from GERDA-Phase II \cite{10} for $^{76}\text{Ge}$, located at the Laboratori Nazionali del Gran Sasso (LNGS), and from KAMLAND-Zen \cite{11} for $^{136}\text{Xe}$, located at the Kamioka Observatory. If the decay process is mediated by the exchange of light Majorana neutrinos, the mismatch between e-flavor neutrino and definite mass neutrinos $\nu_i$ in the Majorana propagator generates a decay amplitude proportional to the effective Majorana neutrino mass 
\begin{equation}
	\label{eq:mbb}
	m_{\beta\beta} \equiv \sum_i  U^2_{ei}\, m_{\nu_i}\,,
\end{equation}
which is a coherent combination of the three neutrino masses. Its determination would then provide a measure of the absolute neutrino mass scale. The effective mass in Eq.(\ref{eq:mbb}) depends on the mixing angles and two relative CP phases of the neutrinos in the $U_\text{PMNS}$ mixing matrix. Assuming that neutrinos are Majorana particles, the present knowledge of mixing angles and neutrino mass differences, from neutrino flavor oscillations, produces the information \cite{12,13} condensed in Figure \ref{fig:1} for the fundamental quantity $m_{\beta\beta}$.

Present experimental limits \cite{14}, indicated in Figure \ref{fig:1}, are approaching the interval of $m_{\beta\beta}$ values predicted for the inverse hierarchy in the neutrino mass spectrum, $\Delta m^2_{13} < 0$. The sign of $\Delta m^2_{13}$ is still an open question in neutrino physics and it is a subjet of current experimental interest \cite{15,16}.

There is an alternative to $0\nu\beta\beta$ by means of the mechanism of neutrinoless double electron capture ($0\nu\text{ECEC}$),
\begin{equation}
	\label{eq:0necec}
	^AZ + 2 e^- \to\, ^A(Z-2)^* \,.
\end{equation}

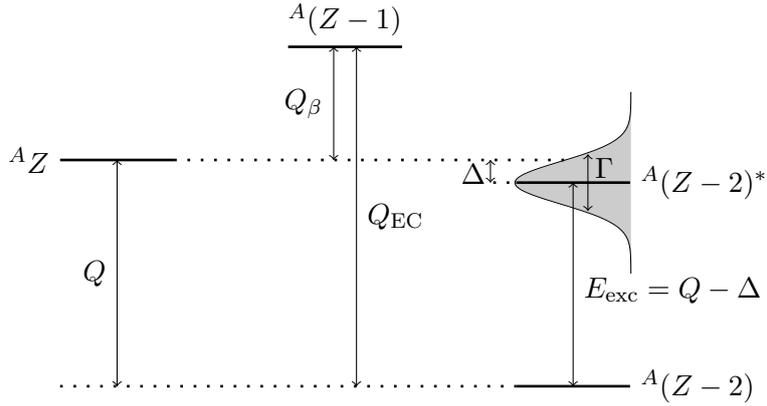
\begin{figure}[t]
	\centering
	\begin{tikzpicture}[line width=1 pt, scale=1.5]
		% Z
		\draw (0,2)--(1,2);
		\node at (0,2)[left]{$ ^AZ$};
		\draw[loosely dotted] (1,2)--(4.5,2);
		% Z-1
		\draw (2,3)--(3,3);
		\node at (2.5,3)[above]{$ ^A(Z-1)$};
		% Z-2
		\draw (4,0)--(5,0);
		\node at (5,0)[right]{$ ^A(Z-2)$};
		\draw[loosely dotted] (0,0)--(4,0);
		% (Z-2)*
		\draw (4,1.8)--(5,1.8);
		\node[rotate=90] at (4.5,1.8){\includegraphics[width=2.5cm, height=1.7cm]{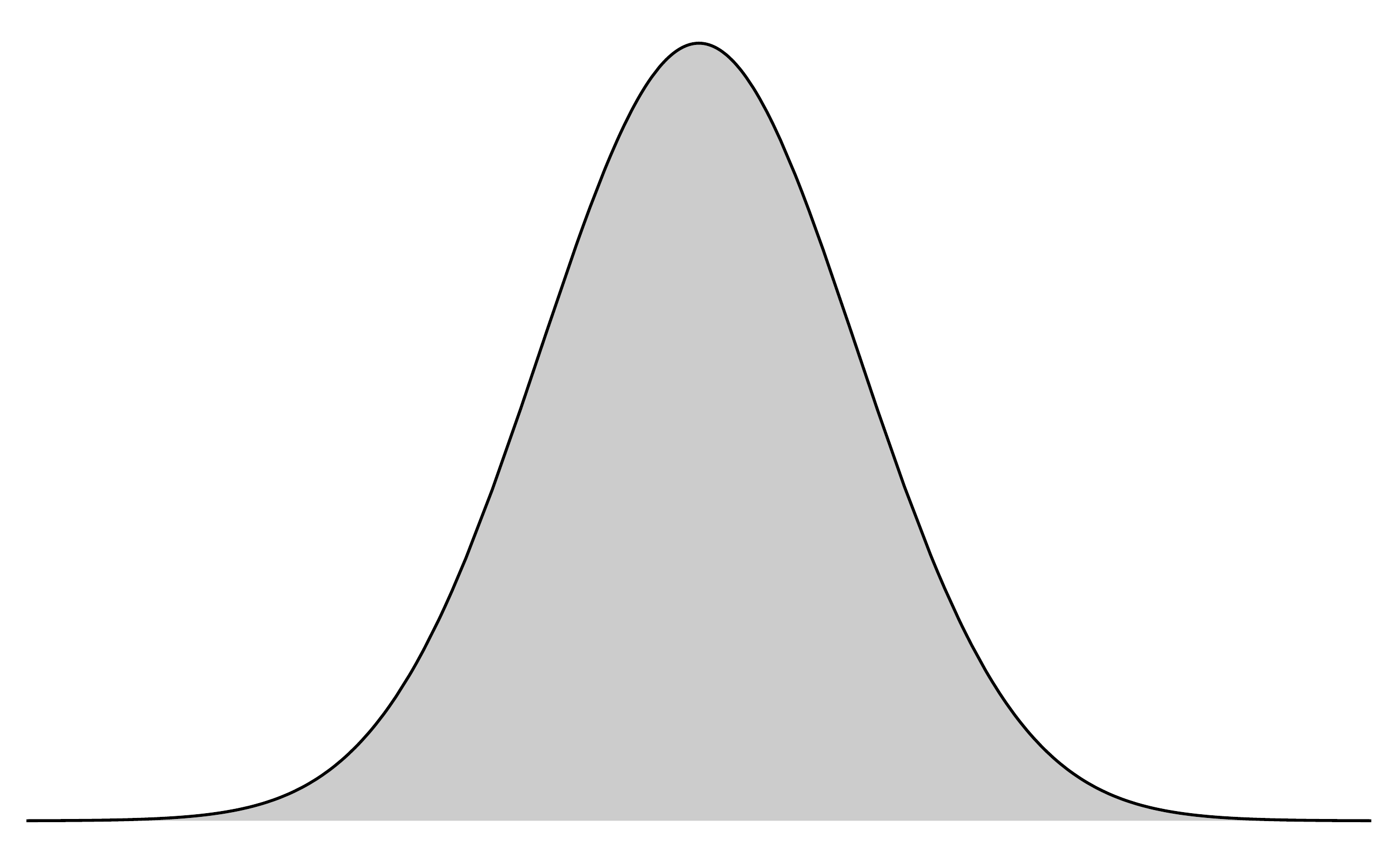}};
		\node at (5,1.8)[right]{$ ^A(Z-2)^*$};
		% Gamma
		\draw[thin, <->] (4.63,1.55)--(4.63,2.05);
		\node at (4.775,1.925){$\Gamma$};
		% Delta 
		\draw[thin, <->] (3.775,1.8)--(3.775,2);
		\draw[loosely dotted] (3.8,1.8)--(4,1.8);
		\node at (3.625,1.875){$\Delta$};
		% Q
		\draw[thin, <->] (0.5,2)--(0.5,0);
		\node at (0.5,1)[left]{$Q$};
		% E_X
		\draw[thin, <->] (4.5,0)--(4.5,1.8);
		\node at (4.5,0.85)[right]{$E_\text{exc} = Q-\Delta$};
		% Qb
		\draw[thin, <->] (2.4,3)--(2.4,2);
		\node at (2.4,2.5)[left]{$Q_\beta$};
		% QEC
		\draw[thin, <->] (2.6,3)--(2.6,0);
		\node at (2.6,1.5)[right]{$Q_\text{EC}$};
	\end{tikzpicture}
	\caption{Schematic representation of the \{$^AZ $, $ ^A(Z-2)^*$\} mixed atomic system.}
	\label{fig:2}
\end{figure}

This is actually a mixing between two states of two different neutral atoms
differing in the total lepton number $L$ by two units, and the same baryonic
number $A$, and not a process conserving energy and momentum in general. The
daughter atom is in an excited state with two electron holes, and its decay
provides the signal for (\ref{eq:0necec}). 
%In Ref.\cite{17} the concept of
%resonant enhancement of $0\nu\text{ECEC}$ was introduced for the exceptional
%circumstance of almost degeneracy between the mother and daughter atomic states
%in (\ref{eq:0necec}). The \emph{almost} matching condition is fulfilled when
%the 2X-ray decay occurs through the tail of the width of the atomic state, as
%shown schematically in Figure \ref{fig:2}.
Ref.~\cite{new18} first pointed out that the monumental coincidence of the
initial energy of the parent atom and that of the intermediate excited atom 
would give rise to a large enhancement of the decay probability.
The concept of
resonant enhancement of $0\nu\text{ECEC}$ was further developed
in~\cite{new19,17} for the exceptional
circumstance of almost degeneracy between the parent and daughter atomic states
in (\ref{eq:0necec}). The \emph{almost} matching condition is fulfilled when
the 2X-ray decay occurs through the tail of the width of the atomic state, as
shown schematically in Figure \ref{fig:2}.
These works stimulated many experimental searches
\cite{18,36,19,20,21,22,23,24,25,26,27} of candidates
when the remarkable trap technique for precision measurements of atomic masses
became available.

%Contrary to $0\nu\beta\beta$, $0\nu\text{ECEC}$  is enhanced for the smallest
%mass differences between the two atomic states in (\ref{eq:0necec}). The work
%of \cite{17} stimulated many experimental searches
%\cite{18,36,19,20,21,22,23,24,25,26,27} of candidates using the trap technique
%for precision measurements of atomic masses, mainly in the last decade. The
%resonant enhancement increases the probability of capture by many orders of
%magnitude, and when the resonance condition is satisfied within a few tens of
%electron-volts, the neutrinoless double electron capture may become competitive
%with neutrinoless double beta decay. In addition, one has to notice that these
%two alternatives present different strategies, different experimental
%signatures and different background conditions; in particular,
%$0\nu\text{ECEC}$ has no irreducible background from the standard
%$2\nu\text{ECEC}$ under the resonance condition. 
%After the improvements in
%measurements of atomic masses, the remaining candidates include
%$^{152}\text{Gd} \to\, ^{152}\text{Sm}$, $^{164} \text{Er} \to\,
%^{164}\text{Dy}$ and $^{180}\text{W} \to\, ^{180}\text{Hf}$ for atomic mixing
%to the daughter atom, with the nucleus in the ground state and having two holes
%in the inner atomic shells.

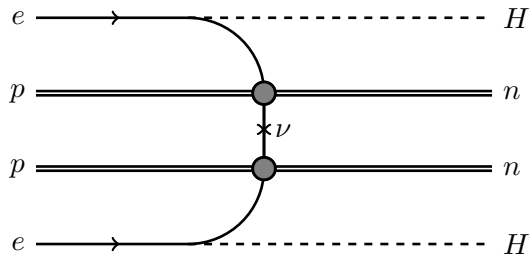
\begin{figure}[b]
	\centering
	\begin{tikzpicture}[line width=1 pt, scale=2]
		% e-H
		\draw[fermion] (0,1.5)--(1,1.5);
		\draw (1,1.5) arc (90:0:0.5);
		\node at (0,1.5)[left]{$e$};
		\draw[dashed] (1,1.5)--(3,1.5);
		\node at (3,1.5)[right]{$H$};
		% p-n
		\draw[double] (0,1)--(3,1);
		\node at (0,1)[left]{$p$};
		\node at (3,1)[right]{$n$};
		% p-n
		\draw[double] (0,0.5)--(3,0.5);
		\node at (0,0.5)[left]{$p$};
		\node at (3,0.5)[right]{$n$};
		% e-H
		\draw[fermion] (0,0)--(1,0);
		\draw (1,0) arc (270:360:0.5);
		\node at (0,0)[left]{$e$};
		\draw[dashed] (1,0)--(3,0);
		\node at (3,0)[right]{$H$};
		% nu
		\draw[fermion] (1.5,0.5)--(1.5,1);
		\draw[fermion] (1.5,1.02)--(1.5,0.52);
		\node at (1.5,0.75)[right]{$\nu$};
		% Blob
		\begin{scope}[shift={(1.5,1)}]
			\draw[fill=black] (0,0) circle (.075cm);
			\draw[fill=black!50] (0,0) circle (.074cm);
		\end{scope}
		% Blob
		\begin{scope}[shift={(1.5,0.5)}]
			\draw[fill=black] (0,0) circle (.075cm);
			\draw[fill=black!50] (0,0) circle (.074cm);
		\end{scope}
	\end{tikzpicture}
	\caption{Feynman diagam for the Majorana mixing amplitude of $^AZ$ atom in the ground state and the 2-hole $^A(Z-2)^*$ state.}
	\label{fig:3}
\end{figure}

The mixing amplitude was calculated in \cite{17} from the diagram in Figure \ref{fig:3} and, in a good approximation, it can be factorized leading to
\begin{equation}
	M_{21} = m_{\beta\beta}^*\,\left( \frac{G_F\cos\theta_C}{\sqrt{2}} \right)^2 \, \left< F_{21} \right> \, \frac{g_A^2}{2\pi} \, M_{0\nu}
	\label{eq:M21}
\end{equation}
%Here $G_\beta = G_F\, cos\theta_C$, where $\theta_C$ is the Cabbibo angle,
%$\left< F_{21} \right>$ depends on the probability amplitude that bound
%electrons are found at the nucleus, $M_{0\nu}$ is the nuclear matrix element,
%$R$ is the nuclear radius,
Here $G_F$ is the Fermi coupling constant, $\theta_C$ is the Cabibbo angle,
$\left< F_{21} \right>$ gives the probability amplitude of finding the two
electrons in the nucleus, $M_{0\nu}$ is the nuclear matrix element, which is of
the order of the inverse nuclear radius, 
$g_A$ is the axial-vector nucleon coupling and the effective Majorana neutrino
mass $m_{\beta\beta}$ appears as the complex conjugate of the expression
(\ref{eq:mbb}) for neutrinoless double beta decay. The experimental activity
in recent years has in turn stimulated the calculation
\cite{28,29,30,31,32,33,34,35} of the nuclear matrix elements for the cases of
interest. 
A list of likely resonant transitions was provided in~\cite{31}, excluding some of
those previously suggested in~\cite{17}.
After the improvements in
measurements of atomic masses, the remaining candidates include
$^{152}\text{Gd} \to\, ^{152}\text{Sm}$, $^{164} \text{Er} \to\,
^{164}\text{Dy}$ and $^{180}\text{W} \to\, ^{180}\text{Hf}$ for atomic mixing
to the daughter atom, with the nucleus in the ground state and having two holes
in the inner atomic shells.
More recent detailed analyses, using the state of the art in nuclear QRPA and IBM
models, agree in the results, showing that the most promising known candidates
are $^{152}\text{Gd}$ and $^{180}\text{W}$.

The case of $^{152}\text{Gd} \to\, ^{152}\text{Sm}$ mixing and decay is
particularly attractive. %For this mixing, 
The values of the relevant
parameters are the experimental $\Delta = M_1-M_2 =
(0.91 \pm 0.18)$~keV \cite{18} for the masses of the parent ``1'' and daughter
``2'' atoms, $\Gamma = 0.023$~keV \cite{36} for the two-hole atomic width and
the $Q$-value of the ground state to ground state transition $Q= (55.70 \pm
0.18)$~keV \cite{18}.
The theoretical mixing is \cite{32,35}
\begin{equation}
	\abs{M_{21}} = 10^{-24}\,
	\left[\frac{\abs{m_{\beta\beta}}}{0.1~\mathrm{eV}}\right]\,
	\mathrm{eV}\,.
\end{equation}
As seen, the optimal resonant enhancement condition is still off
by at least a factor 30, implying a loss of 3 orders of magnitude in the
expected X-ray rate from the parent atom.

In Section 2 we set the formalism for the two-state
atom mixing~\cite{29,31} using a non-normal non-Hermitian Hamiltonian and
obtain the non-orthogonal states with definite time evolution:
one metastable mixed state and one short-lived mixed state.
In Section 3 we develop the natural time history of an
initial parent ground state atom, a history which is governed 
by different time scales of oscillation, lifetime of the short-lived
state and lifetime of the metastable state, thus obtaining the expected
natural populations, at present observable times, of the three 
states involved including the ground state of the daughter atom.
Stimulated by the natural population inversion, we contemplate
in Section 4 the prospects that could be open by both an enhanced rate
of stimulated X-ray emission from the metastable state and the
absorption from the ground state daughter atom.
In Section 5 we present our conclusions and outlook.

%In the following, we develop the time history of the parent atom $^AZ$ since its inception either by Nature or in the laboratory, studying its  natural evolution as a two-state system with the daughter $^A(Z-2)^*$ atom. The purpose is the determination of the relative populations of the three states involved: parent and daughter, as well as the ground state of $^A(Z-2)$. This result allows the identification of appropriate observables, originated either spontaneously or by means of stimulating X-ray beams, as signals of this $\Delta L = 2$ Majorana mixing of atoms.

\section{The evolution Hamiltonian}

In the basis of the $\ket{^AZ\,}$ and $\ket{^A(Z-2)^*}$ states, 
which we'll refer to as 1 and 2, 
the dynamics of this two-state system of interest is governed by the Hamiltonian
\begin{equation}
	\dsH = \dsM - \frac{i}{2}\, \dsGamma
		= \begin{bmatrix}
			M_1 & M_{21}^* \\
			M_{21} & M_2
		\end{bmatrix}
		- \frac{i}{2} \begin{bmatrix}
			0 &0 \\
			0 &\Gamma
		\end{bmatrix}\,,
		\label{eq:H}
\end{equation}
with a Majorana $\Delta L = 2$ mass mixing $M_{21}$ as given by Eq.(\ref{eq:M21}). 
The anti-Hermitian part of this Hamiltonian is due to the instability of $\ket{^A(Z-2)^*}$, 
which de-excites into $\ket{^A(Z-2)_\text{g.s.}}$, external to the two-body system in Eq.(\ref{eq:H}),
emitting its two-hole characteristic X-ray spectrum.
%\textcolor{blue}{The formalism presented to describe the mixing of two-state unstable systems,
%known since Weisskopf-Wigner [ref], was used in Refs.~\cite{29,31} with the
%objective of reproducing the rate induced by the mixing, as given previously in
%Ref.~\cite{17}}.
The appropriate non-Hermitian Hamiltonian formalism for describing the mixing
of a two-state unstable system is known since Weisskopf-Wigner~\cite{ww} and it was
used in many instances. It has been employed~\cite{29,31} with the objective of
reproducing the rate, induced by atom mixing, as previously given in Ref.~\cite{17}.

Besides being non-Hermitian, $\dsH$ is not a normal operator, i.e. $\comm{\dsM}{\dsGamma} \neq 0$. As a consequence, $\dsM$ and $\dsGamma$ are not compatible. The states of definite time evolution, eigenstates of $\dsH$, have complex eigenvalues and are given in non-degenereate perturbation theory \cite{galindo} by

\begin{align}
	\nonumber \ket{\lambda_L} &= \ket{1} + \alpha \ket{2}, \hspace{2cm} \lambda_L \equiv E_L -\frac{i}{2}\, \Gamma_L = M_1 + \abs{\alpha}^2 \left[ \Delta - \frac{i}{2}\, \Gamma \right] , \\
	\label{eq:eigen} \ket{\lambda_S} &= \ket{2} - \beta^* \ket{1}, \hspace{1.9cm} \lambda_S \equiv E_S -\frac{i}{2}\, \Gamma_S = M_2 -\frac{i}{2}\,\Gamma - \abs{\alpha}^2 \left[ \Delta - \frac{i}{2}\, \Gamma \right],
\end{align}
with $\Delta = M_1 - M_2$. As seen in Eq.(\ref{eq:eigen}), $\Gamma_{L,S}$ are
\emph{not} the eigenvalues
of the $\dsGamma$ matrix. The eigenstates are modified at first order in $M_ {21}$,
\begin{equation}
	\label{eq:alpha}
	\alpha = \frac{M_{21}}{\Delta + \frac{i}{2}\,\Gamma}\,,\hspace{2cm} 
	\beta = \frac{M_{21}}{\Delta - \frac{i}{2}\,\Gamma}\,,
\end{equation}
so the ``stationary'' states of the system don't have well-defined atomic properties: both the number of electrons and their atomic properties are a superposition of $Z$ and $Z-2$. Also, these states are \emph{not} orthogonal---their overlap is given by
\begin{equation}
	\braket{\lambda_S}{\lambda_L} = \alpha - \beta =  -i\frac{M_{21}\Gamma}{\Delta^2 + \frac{1}{4}\,\Gamma^2}\,,
\end{equation}
with its non-vanishing value due to the joint presence of the mass mixing $M_{21}$ and the decay width $\Gamma$. 
Notice that Im$(M_{21})$ originates a real overlap.
%Notice that there is a real overlap for the T-violating amplitude Im$(M_{21})$.

As seen in Eq.(\ref{eq:eigen}), the modifications in the corresponding eigenvalues appear at second order in $\abs{M_{21}}$ and they are equidistant with opposite sign. Since these corrections are small, from now on we will use the values
%Even though the correction to $\Gamma$ in $\ket{\lambda_S}$ shown in Eq.(\ref{eq:eigen}) is small, $\ket{\lambda_L}$ acquires a non-zero decay rate to the ground state of $^A(Z-2)$ given by
\begin{align}
	\nonumber E_L &\approx M_1\,, \hspace{2.375cm} E_S \approx M_2\,,\\
	\label{eq:GammaLS}  \Gamma_L &\approx \abs{\alpha}^2\, \Gamma\,,\hspace{2cm} \Gamma_S \approx \Gamma\,.
\end{align}
The only relevant correction at order $\abs{\alpha}^2$ is the one to $\Gamma_L$, since $\ket{1}$ was a stable state---even if it's small, the mixing produces a non-zero decay width.

This result shows that, at leading order, the Majorana mixing becomes observable through $\Gamma_L \propto \abs{\alpha}^2$. The value of $\alpha$ in Eq.(\ref{eq:alpha}) emphasizes the relevance of the condition $\Delta \sim \Gamma$, which produces a resonant enhancement \cite{17} of the effect of the $\Delta L=2$ mass mixing $M_{21}$.

%This result, determined by the value of $\alpha$ in Eq.(\ref{eq:alpha}), emphasizes the relevance of the condition $\Delta \sim \Gamma$, which produces a resonant Enhancement \cite{17} of the effect of the $\Delta L=2$ mass mixing $M_{21}$ in the mixing $\alpha,\beta$ of the physical states.
%From Eq.(\ref{eq:alpha}) on the mixing states, and particularly from Eq.(\ref{eq:GammaLS}) for the observable $\Gamma$, we realize the relevance of the resonance condition $\Delta \sim \Gamma$ in order to enhance the effect of the $\Delta L = 2$ mass mixing $M_{21}$.

\section{Natural time history for initial $\ket{^AZ\,}$}

As seen in Eq.(\ref{eq:eigen}), the states $\ket{^AZ\,}$ and $\ket{^A(Z-2)^*}$
are not the stationary states of the system. For an initially prepared
$\ket{^AZ\,}$, the time history is far from trivial and the appropriate
language to describe the system short times after is that of \emph{atom
oscillations}~\cite{29} between $\ket{^AZ\,}$  and $\ket{^A(Z-2)^*}$ due to the interference of the amplitudes through $\ket{\lambda_S}$ and $\ket{\lambda_L}$ in the time evolution. The time-evolved $\ket{^AZ\,}$ state becomes
\begin{equation}
	\label{eq:slits}
	\ket{^AZ(t)} = e^{-i\lambda_L t}\ket{\lambda_L} - \alpha\, e^{-i\lambda_S t}\ket{\lambda_S}
\end{equation}
and the appearance probability is then given by
\begin{equation}
	\label{eq:t1}
	\abs{\braket{^A(Z-2)^*}{^AZ(t)}}^2 = \abs{\alpha}^2 \left\{ 1 + e^{-\Gamma t} - 2e^{-\frac{1}{2} \Gamma t}\cos(\Delta \cdot t) \right\}\,,
\end{equation}
with an oscillation angular frequency $\abs{\Delta}$. The characteristic oscillation time 
$\tau_\text{osc} = 2\pi~\abs{\Delta}^{-1}$ is the shortest time scale in this system.
For $t \ll \tau_\text{osc}$, one has
\begin{equation}
	\abs{\braket{^A(Z-2)^*}{^AZ(t)}}^2 \approx \abs{M_{21}}^2\, t^2
\end{equation}
induced by the mass mixing. 

The next shortest characteristic time in this system is the decay time $\tau_S = \Gamma^{-1}$, associated to the $\ket{\lambda_S}$ state. For $\tau_\text{osc} \ll t \ll \tau_S$, the only change with respect to Eq.(\ref{eq:t1}) is that the interference region disappears, and the two slits $\ket{\lambda_L}$ and $\ket{\lambda_S}$ in (\ref{eq:slits}) contribute incoherently,
\begin{equation}
	\abs{\braket{^A(Z-2)^*}{^AZ(t)}}^2 \approx \abs{\alpha}^2\,\left( 2-\Gamma t \right)\,.
\end{equation}

\begin{figure}[t]
%	\centering
	\begin{tikzpicture}[scale=1.65]
	%Definitions
	\def\a{0.1}	%hline
	\def\l{8.5}	%length
	\def\to{1}	%tau_osc
	\def\ts{1.5}	%tau_s
	\def\dt{2}	%exp resol
	\def\t{4.5}	%t0
	\def\Dtx{0}%{0.075}
	\def\tl{7}	%tau_L
	\def\lab{1.125}	%Labels size
	
	%Arrow:
	\draw[-stealth', line width=2.5pt] (0,0)--(\l,0);
	%\node[scale = \lab] at (\a,-{0.99*\l})[right]{$t$};
	
	%Creation
	\draw[line width = 2.5pt] (0,-\a)--(0,\a);
	\node[rotate = 45, scale = \lab] at (0,\a)[above right]{$^AZ$ produced};
	
	%tau_osc
	\draw[line width = 2.5pt] (\to,-\a)--(\to,\a);
	\node[rotate = 45, scale = \lab] at (\to,\a)[above right]{$\tau_\text{osc} \sim 10^{-18}$ s};
	
	%tau_s
	\draw[line width = 2.5pt] (\ts,-\a)--(\ts,\a);
	\node[rotate = 45, scale = \lab] at (\ts,\a)[above right]{$\tau_S\sim 10^{-17}$ s};

	%delta t
	\draw[line width = 2.5pt] (\dt,-\a)--(\dt,\a);
	\node[rotate = 45, scale = \lab] at (\dt,\a)[above right]{$\delta t\gtrsim 10^{-15}$ s};

	%t0
	\draw[line width = 2.5pt] (\t,-\a)--(\t,\a);
	\node[rotate = 45, scale = \lab] at (\t,\a)[above right]{$t_0 <
		t_{\pmb{\oplus}} \sim 10^{17}$ s $\sim 10^{9}$ yr};
	
	%Delta t
	\draw[->, line width = 2pt, blue] (\t+0.3,\Dtx)--(\t, \Dtx);
	\draw[->, line width = 2pt, blue] (\t+0.3,\Dtx)--(\t+0.6,\Dtx);
	\node[scale = \lab] at (\t+0.3,\Dtx)[above]{\textcolor{blue}{$\Delta t$}};
	
	%tau_L
	\draw[line width = 2.5pt] (\tl,-\a)--(\tl,\a);
	\node[rotate = 45, scale = \lab] at (\tl,\a)[above right]{$\tau_L\sim
		10^{36}$ s $\sim 10^{29}$ yr};
\end{tikzpicture}
\caption{Timeline of the mixed atomic system.
	The $\tau_\text{osc},\, \tau_S$ and $\tau_L$ values are for $^{152}\text{Gd}$$\to$$^{152}\text{Sm}$,
	whereas $\delta t$ is a typical time resolution, $t_0$ is the age of the sample ore,
	$t_{\pmb{\oplus}}$ is the age of the Earth and $\Delta t$ is the observation time in an actual experiment.}
\label{fig:timeline}
\end{figure}
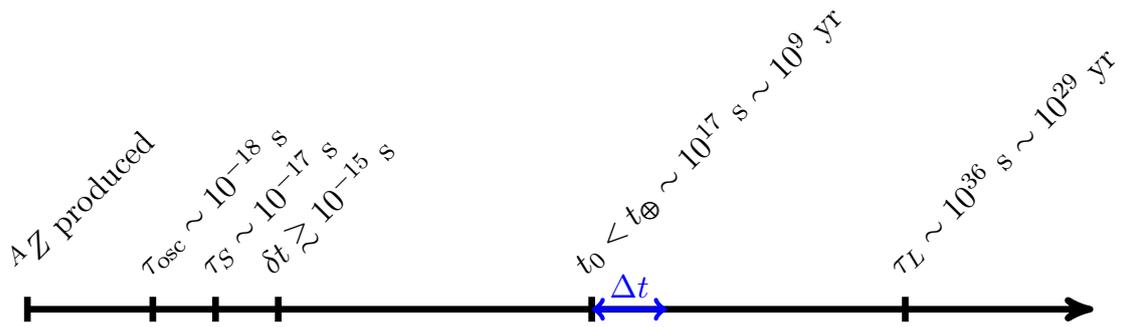

For $t\gg \tau_S$, the contribution of $\ket{\lambda_S}$ disappears and the appearance probability simply becomes
\begin{equation}
	\abs{\braket{^A(Z-2)^*}{^AZ(t)}}^2 = \abs{\alpha}^2\,.
\end{equation}
In other words, the initially prepared $\ket{^AZ\,}$ state evolves towards the stationary metastable state $\ket{\lambda_L}$,
\begin{equation}
	\label{eq:tl}
	\ket{^AZ(t)} \to e^{-i\lambda_L t}\ket{\lambda_L}\,,
\end{equation}
with the long lifetime $\tau_L = \Gamma_L^{-1}$ from Eq.(\ref{eq:GammaLS}). For a realistic time resolution $\delta t$ in an actual experiment, this regime is the interesting one, with the behavior in Eq.(\ref{eq:tl}). As shown in Figure \ref{fig:timeline}, the different time scales involved in this problem are thus
\begin{equation}
	\boxed{\hspace{0.5cm}\tau_\text{osc} \ll \tau_S \ll \delta t \ll t \ll \tau_L \hspace{0.5cm}\begin{aligned}\mbox{}\\ \mbox{} \end{aligned}}
\end{equation}
where $t$ refers to the elapsed time since the production of $^AZ$, either by nature or in the lab---given the smallness of the mixing, the metastability of the state (\ref{eq:tl}) is valid even for cosmological times. Therefore, for any time between the two scales $\tau_S$ and $\tau_L$, the populations of the three states involved are given by the probabilities\\
\begin{minipage}{0.75\textwidth}
\vspace{-0.8cm}\begin{align*}
	\mbox{}\\
	\hspace{3cm}\boxed{ \hspace{1cm} \tau_S \ll t \ll \tau_L \hspace{0.5cm} \Longrightarrow \hspace{0.5cm}
	\left\{\begin{aligned}
		P_L(t) &\approx 1-\Gamma_L\, t\\
		P_S(t) &\approx 0\\
		P_\text{g.s.}(t) &\approx \abs{\alpha}^2\, \Gamma\, t
	\end{aligned} \right. \hspace{1cm} \begin{aligned} \mbox{}\\\mbox{}\\\mbox{}\\\mbox{} \end{aligned}}\\
	\mbox{} 
\end{align*}
\end{minipage}
\hfill
\begin{minipage}{0.15\textwidth}
\vspace{-0.88cm}
\begin{subequations}
	\label{eqs:P}
	\begin{align}
		\mbox{} \label{eq:PL}\\
		\mbox{} \label{eq:PS}\\
		\mbox{} \label{eq:Pgs}
	\end{align}
\end{subequations}
\end{minipage} \\
where $P_\text{g.s.}(t)$ refers to the population of the ground state of the $^A(Z-2)$ atom after the decay of the unstable ``stationary'' state $\ket{\lambda_S}$ in (\ref{eq:slits}), with rate $\Gamma$. No matter whether $t$ refers to laboratory or cosmological times, the linear approximation in $t$ is excellent.

With this spontaneous evolution of the system, an experiment beginning its measurements 
a time $t_0$ after the $^AZ$ was produced will probe the three-level system 
with relative populations $P_L \approx 100\%,\, P_S \approx 0,\,  P_\text{g.s.}\approx \abs{\alpha}^2\Gamma\, t_0$. 
We discover two methods, involving the third state beyond the mixed states\footnote{
One may wonder whether there is, for $\Delta>0$,
a spontaneous emission of lower energy X-rays from $\ket{\lambda_L}$ to $\ket{\lambda_S}$
leading to a regeneration of the short-lived mixed state.
Observing in (\ref{eq:eigen}) the atom mixing of these states,
the dynamics of this process would be that of the Compton amplitudes
for the $Z$ and $(Z-2)^*$ atoms,
whereas the kinematics corresponds to two-photon emission instead of scattering.
At these intermediate energies between atomic and nuclear physics,
the Compton amplitude $T_{2\gamma}$ can be taken to be 
an incoherent sum of the electron contributions \cite{rosa-clot, rosa-clot2},
$$ T_{2\gamma} = \alpha\, T_{2\gamma}^{Z-2} - \beta\, T_{2\gamma}^{Z} 
\hspace{1cm} ; \hspace{1cm}
T_{2\gamma}^Z = \frac{Z\, e^2}{m}\,(\bm{\epsilon}^{\prime\, *} \cdot \bm{\epsilon}^*) $$
where $m$ is the electron mass and $\bm{\epsilon}$ the polarization vectors of the photons.
A straightforward calculation of the rate for this e.m. 
$\ket{\lambda_L}\to\ket{\lambda_S}$ transition,
when compared to the transition to the daughter atom ground state,
gives a branching ratio of the order $10^{-7}$.},
to be sensitive to the resonant Majorana mixing of atoms:
\begin{itemize}
	\item {\bf Spontaneous emission from the metastable state to the daughter
		atom ground state.} The population in the upper level
		$\ket{\lambda_L}$, as shown in Eq.(\ref{eq:PL}), decreases with time as
		$P_L(\Delta t) \approx 1 - \Gamma_L\, \Delta t$, where $\Delta t = t -
		t_0$, due to the decay of the metastable ``stationary'' state
		$\ket{\lambda_L}$ to $\ket{^A(Z-2)_\text{g.s.}}$. This process is
		associated to the spontaneous emission of X-rays with a rate
		$\Gamma_L$, considered in the literature after the concept of resonant
		mixing was introduced in Ref.~\cite{17}. For one mole of $^{152}$Gd, the
		X-ray emission rate would be of order $10^{-12}$~s$^{-1}\sim
		10^{-5}$~yr$^{-1}$. 
		%The initial state at observable times being
		%$\ket{\lambda_L}$, its unique
		%signature is the total energy of the two-hole X-ray radiation,
		The initial state in the transition at observable times, being 
		$\ket{\lambda_L}$, tells us that the total energy of the two-hole
		X-ray radiation is 
		displaced by $\Delta$ with respect to the
		characteristic $\ket{^A(Z-2)^*}\to \, \ket{^A(Z-2)_\text{g.s.}}$ X-ray
		spectrum, i.e. its energy release is the $Q$-value between the two
		atoms in their ground states (as seen in Figure \ref{fig:2}).

	\item {\bf Daughter atom population.} The presence of the daughter atom
		in the parent ores (see
		Eq.\ref{eq:Pgs}), can be probed e.g. by geochemical methods. For one mole of the
		nominally stable $^{152}$Gd isotope produced at the time of the Earth
		formation, the values in Figure \ref{fig:timeline} would predict an
		accumulated number of order $10^4$ $^{152}$Sm atoms.
		This observable could be of interest for cosmological times
		$t_0$ since, contrary to $\beta\beta$-decay, in the ECEC case there is
		no irreducible background from a $2\nu$ channel for a resonant atom
		mixing.

\end{itemize}

We would like to emphasize that, even though $\Gamma_L = \abs{\alpha}^2 \Gamma$
ensures the probability conservation, an interpretation of Eqs.(\ref{eq:PL})
and (\ref{eq:Pgs}) in different terms is of interest. On the one hand,
$\Gamma_L$ is the rate for the decay of $\ket{\lambda_L}$ at any time $t$,
which accounts for the observable {\bf 1}. On the other hand, the population of
the daughter atom in the ground state is obtained from the mixing probability
leading to $\ket{\lambda_S}$ in Eq.(\ref{eq:slits}) at all times, given by
$\abs{\alpha}^2$, times its decay rate to the ground state, $\Gamma_S =
\Gamma$. This mixing$\times$Decay temporal evolution explains the non-zero
population of $^A(Z-2)_\text{g.s.}$, producing the second observable.

\section{Stimulated transitions}

\subsection{Emission from $\ket{\lambda_L}$}
A careful reading of Eqs.(\ref{eqs:P}) shows that the metastable state
$\ket{\lambda_L}$ and the ground state $\ket{^A(Z-2)_\text{g.s.}}$ have a
\emph{natural population inversion}, with an overwhelming abundance of the
long-lifetime upper level of the system. This result suggests the exploitation
of the bosonic properties of the X-radiation, used as a signal of the Majorana
mixing in this problem, and considering the external action of an X-ray beam to
stimulate the emission from the metastable level to the ground state of this
atomic system.

Stimulated radiation for the emission from $\ket{\lambda_L}$ to the ground state 
$\ket{^A(Z-2)_\text{g.s.}}$ could then enhance the rate and we present an estimate 
of the gain which could be envisaged in future facilities of X-ray beams.
A setup with an incident pulsed beam allows the observation 
of low rate events in directions outside the beam direction
and the control of background conditions in the absence of the beam.
Therefore, one discovers a third observable

\begin{itemize}
	\item {\bf\boldmath Stimulated emission from $\ket{\lambda_L}$ to
		$\ket{^A(Z-2)_\text{g.s.}}$.} The natural population inversion between
		the ground state and the metastable ``stationary'' state
		$\ket{\lambda_L}$ gives raise to the possibility of stimulating the
		decay $\ket{\lambda_L} \to \ket{^A(Z-2)_\text{g.s.}}$. The experimental
		signature of this process would be the emission of X-rays with total
		energy equal to the $Q$-value of the process, just like in the first
		observable of spontaneous emission.
		%The advantage of this observable is clear: the adequate beam may
		%enhance the stimulated emission rate by many orders of magnitude.
\end{itemize}

For the emission between the two levels $\ket{\lambda_L}\to\ket{^A(Z-2)_\text{g.s.}}$
of radiation with angular frequency $\omega$,
stimulated radiation is described in terms of the Einstein coefficients \cite{einstein}
with an induced rate\footnote{For the sake of clarity, throughout this
discussion we keep all $\hbar$ and $c$ factors.}
\begin{equation}
	\dv{N_L}{t} = -\frac{\pi^2 c^3}{\hbar \omega^3}\,\rho_\omega\, \Gamma_L N_L\,,
\end{equation}
where $N_L$ is the population of the upper metastable $\ket{\lambda_L}$ level, 
$\Gamma_L$ its width and 
$\rho_\omega$ is the energy density of the beam per unit of angular frequency, i.e.
\begin{equation}
	\rho_\omega = \frac{\dd E}{c\dd{t}\dd{S}\dd{\omega}}\,.
	\label{eq:rhoomega}
\end{equation}

Therefore, this observable is enhanced with respect to the first one by a gain factor
\begin{equation}
	G = \frac{\pi^2\, c^3}{\hbar\, \omega^3}\, \rho_\omega\,,
\end{equation}
which is the ratio between the stimulated and spontaneous emission rates.

In order to produce a sizable gain, one should devise a setup with as large a $\rho_\omega$ as possible.
The transition energy of this system is of order tens of keV,
so a high-luminosity X-ray beam is mandatory.
Such high-energy beams are produced at free-electron laser (FEL) facilities, 
through a kind of laser consisting of very-high-speed electrons moving freely through a magnetic structure.
Free-electron lasers are tunable and have the widest frequency range of any laser type, 
currently ranging in wavelength from microwaves, through terahertz radiation
and infrared, to the visible spectrum, ultraviolet, and X-ray. The highest
frequencies are obtained in XFEL facilities
like the running SLAC Linac Coherent Light Source (LCLS) and the commissioned
European XFEL (EXFEL) at DESY.

The determination of the gain factor one could achieve in these facilities is clearer after rewriting
the spectral energy density (\ref{eq:rhoomega}) in terms of beam parameters,
\begin{equation}
	\rho_\omega = \frac{\hbar}{c}\, \frac{\dd N}{\dd{t}\dd{S}}\, \left[ \frac{\dd\omega}{\omega} \right]^{-1}\,.
\end{equation}
Taking $\dd N/\dd t$ as the number of photons per pulse duration,
$\dd S$ as the beam section and $\dd\omega/\omega$ as the full width half
maximum (FWHM) spectrum width, one finds the gain factor
\begin{equation}
	\boxed{\hspace{0.5cm}G = \hbar\, (\hbar c)^2\, \frac{\pi^2}{(\hbar \omega)^3}\, \frac{\dd N}{\dd{t}\dd{S}}\,
	\left[ \frac{\dd \omega}{\omega} \right]^{-1}\hspace{0.5cm}\begin{aligned}\mbox{}\\ \mbox{} \end{aligned} }
	\label{eq:Gain}
\end{equation}
to be written in terms of clearly defined beam properties, where $\dd N/\dd t\dd S$ is the luminosity $\cal L$ of the beam.

At EXFEL, a sound simulation of the conditions of the machine \cite{altarelli} gives,
for typical energies of tens of keV,
the expected number of photons per pulse duration $dN/dt = 10^{10}$~fs$^{-1}$
and the spectral width $\dd\omega/\omega = 1.12\times 10^{-3}$.
Nanofocusing of this X-ray FELs has been contemplated \cite{gain};
using a beam spot of the order of 100~nm would lead to a gain factor
from (\ref{eq:Gain}) of $G\sim 100$. 
%Applying this beam on a target of
%one mole of $^{152}$Gd would provide an X-ray emission rate of order
%$10^{-10}$~s$^{-1}\sim 10^{-3}$~yr$^{-1}$.
The continuous interaction of these X-rays with a mole
of $^{152}$Gd atoms would provide a stimulated rate of
order $10^{-10}$~s$^{-1}\sim 10^{-3}$~yr$^{-1}$.

%\textcolor{blue}{
%Notice that this number of events would only apply to an experiment where the whole target is
%constantly irradiated. 
One should notice that this rate of events assumes constant irradiation of the
whole target.
The straightforward setup of a cylindrical target
alongside the pulsating X-ray beam presents different issues. Most notoriously,
pulsed beams limit the enhancing time to a fraction of the running time.
EXFEL manages to produce $2.7\times 10^4$ pulses per second, so that the
fraction of effective time is of order $10^{-9}$; LCLS-II expects to produce
pulses at $1$~MHz, increasing this number by two orders of magnitude, but
still far away from a promising factor. Furthermore, radiation of these energies
has an attenuation length in Gd of tens of microns, limiting the amount of
material one could use to a fraction of a mole.
This setup has the general drawback that the
high energy density effect associated with the small beam spot size
is lost when considering the small interaction volume.
%}

%\textcolor{blue}{
%This attenuation of the beam exists due to its ionizing the sample, which in
%turn leads to its heating. 
The attenuation of the beam is associated to its interaction with the sample,
which is dominated by the photoelectric effect and, to a lesser extent, inelastic 
Compton scattering, leading to ionization. Successive interactions of the secondary electrons
will heat the material. 
A recent simulation \cite{Heating} of this effect under realistic
experimental conditions on a cylindrical target, assuming the extreme limit
that the whole absorption power is converted into heating power, shows that a
temperature of about $700^\circ$C is reached for an incoming beam of spot size $100$~nm and
an average of $10^{14}$~X-ray photons/s. Since this temperature
is proportional to the flux, in all high-flux experiments like the one
contemplated in this work, the small-interaction-volume target is actually destroyed.
The design of a macroscopic sample whith a very large number of thermally isolated 
micro-targets,
built on a plane in transversal motion synchronized with the pulse frequency of
the beam, is a subject of current interest \cite{Microtarget}.
The use of this approach in order to stimulate 
the $\Delta L=2$ emission rate 
should be explored after a suitable candidate is found.
On the other hand, the limiting factors in the expected integrated rate of events
also suggest an alternative
ingenuity program more in the line of micro-particles inserted into a dreamed
X-ray resonant cavity.
%Therefore, after a suitable candidate is found, an 
%experiment of this kind should take into account the need of a cooling system. 
%}

%\textcolor{blue}{A different approach may be more appealing.}
%On the other hand, 
\vspace{6pt}
\subsection{Absorption from $\ket{A(Z-2)_\mathrm{g.s.}}$}
A different observable may also be considered.
The existing population of the daughter atom in its ground state is,
by itself, a signal of the atom Majorana mixing, as discussed in the previous Section as
a relic of the previous history with an initial parent atom.
In addition, this population can be identified by using an intense photon beam, leading
to the characteristic absorption spectrum of the daughter atom and its subsequent decay to the ground state.
\begin{itemize}
	\item \textbf{Stimulated absorption spectrum of the daughter atom.} In the
		presence of a light beam, the daughter atom population would absorb
		those characteristic frequencies corresponding to its energy levels, which would then
		de-excite emitting light of the same frequency.	In the case of the one mole
		$^{152}$Gd ore that we mentioned in the previous section, all $10^4$ Sm
		atoms could be easily excited to any of its $\sim 1$~eV levels using a standard pulsed laser of order
		$100$~fs pulse duration, with a mean power of $5$~W and a pulse rate of 
		$100$~MHz.
\end{itemize}
Notice that these numbers imply, for a laser with FWHM spot size $\sim
40$~$\mu$m, an absorption rate
\begin{equation}
	\left. \dv{N_\mathrm{g.s.}}{t} \right|_\mathrm{abs} = -60\%\,
		N_\mathrm{g.s.}\, \left[\frac{100~\mathrm{ns}}{\tau}\right] \;\mathrm{fs}^{-1}\,.
\end{equation}
Since Sm levels have lifetimes between $10-1000$~ns \cite{SmLifes}, one expects to excite them
all during the $100$~fs pulse.
Disentangling the parent and daughter lines should not be difficult---the
relatively small number of atomic absorption lines (compared to atomic emission
lines) and their narrow width (a few pm) make spectral overlap rare, not being
expected between $Z$ and $(Z-2)$ atoms.

It is worth noting from the results of this section that the bosonic nature of
the atomic radiation is a property that can help in getting observable rates of
the atom Majorana mixing, including the stimulated X-ray emission from the
parent atom as well as the detection of the presence of the daughter atoms by
means of its characteristic absorption lines. The actual values correspond to
the specific case of $^{152}$Gd~$\to$~$^{152}$Sm, which is still off the resonance 
condition by at least a factor 30, implying a factor $10^3$ in the rates.

\section{Conclusions}

Neutrinoless double electron capture in atoms is a quantum mixing
mechanism between the neutral atoms $^AZ$ and $^A(Z-2)^*$ with two electron holes.
It becomes allowed for Majorana neutrino mediation responsible 
of this $\Delta L = 2$ transition.
This Majorana mixing leads to the 
X-ray de-excitation of the $\ket{^A(Z-2)^*}$ daughter atomic state which,
under the resonance condition, has no Standard Model background
from the two-neutrino decay.

The intense experimental activity looking for atomic candidates
satisfying the resonance condition by means of precise measurements of atomic masses, 
thanks to the trapping technique, has already led to
a few cases of remarkable enhancement effects and there is still room 
for additional adjustements of the resonance condition. With this
situation, it is important to understand the complete time evolution
of an atomic state since its inception and whether one can find,
from this information, different signals of the Majorana mixing, including the
possible enhancement due to the bosonic nature of atomic transition radiation.
These points have been addressed in this work.

The effective Hamiltonian for the two mixed atomic states
leads to definite non-orthogonal states of mass and lifetime, each of them violating global lepton number,
one being metastable with long lifetime and the other having a short lifetime.
For an initial atomic state there are time periods of atom oscillations, 
with frequency the mass difference, and the decay of the short lived state, which are not observable
for present time resolutions.
For observable times, the system of the two atoms has three relevant states for discussing transitions: 
one highly populated state with long lifetime, 
one empty state with short lifetime 
and the ground state of the daughter atom with a small population as a result of the past history. 
As a consequence, this is a case of natural population inversion suggesting the possibility 
of stimulated radiation transitions besides the natural spontaneous X-ray emission. 

The gain factor for stimulated emission due to a present radiation 
energy density per unit frequency has been adapted to the case of an X-ray beam in terms
of conventional parameters like its luminosity, the energy and the
spectrum width. Using simulated previsions of the now commissioned European XFEL facility,
with a beam spot size of $100$ nm, we obtain an
expected gain of $100$ in the X-ray emission rate from the metastable
long-lived state.
This substantial gain by stimulating the X-ray emission
from $\ket{\lambda_L}$ to $\ket{^A(Z-2)_\mathrm{g.s.}}$ for interacting X-ray photons with
atoms is, however, not exploited in a straightforward setup of a
single pulsed beam directed towards a cylindrical target. The limiting factors
of the pulse frequency and the small interaction volume are
suppressing that ideal benefit for the integrated number of events.

The small population of the ground state daughter atom at observable times
would be by itself a proof of the atomic Majorana mixing, given the absence of
the standard two neutrino decay of the parent atom. Besides other geochemical
methods, absorption rates and the subsequent emission by an intense photon beam
for the daughter atom, due
to its non-vanishing population of the ground state, can be clearly
contemplated.

The results obtained in this work demonstrate that the knowledge of the time
history of a nominally stable atom since its inception can be a source of
inspiration to find appropriate observables in the search of 
evidence for $\Delta L = 2$ double electron capture. The natural population
inversion at observable times suggests stimulating the X-ray emission 
with gain factors which could become significant for appropriate setups. 
On the other hand, the presence of the daughter atom ground state can be signaled by
looking for its characteristic absorption spectrum. Taking into account the
ongoing searches for new isotope candidates with a better fulfillment of the
resonance condition, it remains to be seen whether these processes, with the
ideas on stimulating the transitions, could become actual alternatives in the
quest for the Dirac/Majorana nature of neutrinos.

%The small population of the ground state daughter atom at
%observable times would be by itself a proof of the atomic Majorana
%mixing, given the absence of the standard two neutrino decay of the parent atom.
%Besides other geochemical methods, absorption rates
%---and the subsequent emission---
%by an intense photon beam for the daughter atom,
%due to its non-vanishing population of the ground state,
%can be clearly contemplated.
%
%\textcolor{red}{The actual results obtained in this work demonstrate that the gain factors which
%could be obtained for double electron capture signals, by using the strategy of
%stimulating the relevant  transitions, are significant. Taking into account the
%ongoing searches for new atomic candidates with a better fulfillment of the resonance
%condition, these processes could become realistic alternatives in the quest for the
%Dirac/Majorana nature of neutrinos.}
%
%\textcolor{blue}{The actual results obtained in this work demonstrate that the gain factors which
%could be obtained for double electron capture signals, by using the strategy of
%stimulating the relevant  transitions, are significant. However, we also show
%that that the straightforward setup of a cylindrical sample along a pulsed beam
%does not exploit the benefits of the calculated gain factors. More effort
%should be devoted to find out whether experiments based on the ideas here
%presented could become realistic alternatives in the quest for the
%Dirac/Majorana nature of neutrinos.}

\acknowledgments

The authors would like to acknowledge scientific 
discussions with, and advice from, Massimo Altarelli,
Michael Block, Albert Ferrando, Gaston Garcia and Pablo Villanueva.
This research has been supported by MINECO Project 
FPA 2014-54459-P, Generalitat Valenciana Project GV 
PROMETEO II 2013-017 and  Severo  Ochoa  Excellence  
Centre  Project  SEV  2014-0398. A.S. acknowledges the 
MECD support through the FPU14/04678 grant.

\end{document}